# Real-Time Modeling of Skyrmion Dynamics in Arbitrary 2D Spatially Dependent Pinning Potential Landscapes

Simon M. Fröhlich[1,*], Tobias Sparmann[1,*], Maarten A. Brems[1], Jan Rothörl[1], Fabian Kammerbauer[1], Klaus Raab[1], Sachin Krishnia[1], Mathias Kläui[1, §], Peter Virnau[1, †]

[1]Institute of Physics, Johannes Gutenberg University Mainz, 55099 Mainz, Germany

[*]Contributed equally

[§]klaeui@uni-mainz.de

[†]virnau@uni-mainz.de

## Abstract

Non-flat energy landscapes leading to localized pinning of skyrmions pose an inherent and unavoidable challenge for studies of fundamental 2D spin structure dynamics as well as applications. Accounting for pinning is a key requirement for predictive modeling of skyrmion systems, as it impacts the systems' dynamics and introduces randomizing effects. In this article, we use magneto-optical Kerr microscopy to image skyrmions in a magnetic thin film system in real time and analyze their hopping dynamics within the non-flat energy landscape. To achieve a fully quantitative model, we utilize skyrmion diffusion and dwell times at pinning sites to extrapolate the pinning energy landscape into regions that cannot be sampled on reasonable experimental time scales. For evaluation with a coarse-grained Thiele model, we perform long-time measurements of skyrmion diffusion and then develop a two-step procedure to determine simulation parameters by comparing them to experimentally accessible behavior. This provides a direct conversion between simulation and experimental units, which is the missing key step that has previously prevented quantitative quasiparticle modeling. We demonstrate the predictive power of our approach by measuring the experimentally unexplored density dependence of skyrmion diffusion and show that it is in excellent agreement with simulation predictions. Our technique thus enables quantitative skyrmion simulations on experimental time and length scales, allowing for predictive in-silico prototyping of skyrmion devices.

# Introduction

Magnetic skyrmions are topologically stabilized two-dimensional whirls of magnetization[1,2] which occur in a variety of systems ranging from thin film multilayers to bulk systems[1,3–5]. They can be displaced deterministically by using effective external forces, such as spin polarized current- or magnetic field gradients [6–9], and in some systems, additionally, they exhibit sizable thermal diffusion[10–12]. Both phenomena make skyrmions promising candidates as information carriers in racetrack memory[13], as well as non-conventional computing devices[4,14,15], implementing paradigms such as low-power Brownian[16–19] or probabilistic computing[10].

The motion of skyrmions, whether induced by currents or thermal fluctuations, is strongly influenced by the spatially inhomogeneous energy landscape of the sample[20–24]. Skyrmions and domain walls in general have been observed to exhibit hopping-type motion[10,20,25] due to pinning, with externally driven skyrmions typically exhibiting pronounced motion in the creep regime[7,26–29]. Experiments have shown that skyrmions—particularly those with narrow domain walls relative to their radius, which are stabilized predominantly by dipolar fields—are typically pinned at the domain wall position[20]. However, detailed and complete quantifications of the pinning energy landscape have been limited to strongly confined or one-dimensional systems[30]. In contrast, studies of global energy landscapes—essential for understanding and modeling devices for computing, as well as for studying the formation of 2D skyrmion lattices[31–34]—have largely been qualitative or sufficiently quantify only small subspaces of the overall sample structure[20,31,35]. This limitation stems from the impracticality or even impossibility of protracted measurement times required to collect sufficient statistics for describing the full energy landscape of large experimental systems. This especially holds for spatial resolutions sufficient to describe dynamics within individual pinning sites as the corresponding potential varies on scales smaller than the skyrmion size due to the domain wall-related origin of pinning effects[20,35]. The spatially inhomogeneous energy landscapes result from the skyrmions' interactions with material defects, which are unavoidable in state-of-the-art systems, yet are often essential for competitive realizations of non-conventional computing devices[14]. In ultra-low-power Brownian computing, for instance, pinning effects compete with thermal activity[14,16–19] and often with current-induced forces on similar energy scales[14,16,19]. Additionally, there is a tradeoff between system complexity and diffusion with varying skyrmion number[10,36,37]. Thus, a detailed understanding and quantitatively predictive modeling of pinning and density effects in real experimental systems is key to the advancement of these technologies.

To computationally model the effects of pinning for the dynamics, previous studies have employed atomistic and micromagnetic simulations, incorporating spatial variations in magnetic properties, which has shown to be reliable in small skyrmion systems and on short timescales[20,22,38]. However, simulating pinning effects for large skyrmions or multi-skyrmion systems on experimentally relevant time- and length scales suffers from prohibitive computational costs[20]. Coarse-grained Thiele model simulations, which treat skyrmions as quasiparticles, offer a significantly faster alternative[39,40]. These models have been successful in describing the static behavior of skyrmion systems, but have been limited to qualitative approaches when it comes to dynamic studies[39,41,42], with quantitative predictions available only for specifically designed geometrically confined systems[30]. This limitation arises from the lack of reliable methods to convert simulation timescales to experimental timescales, which is directly related to determining the effective system damping, and to accurately represent the energy landscape within simulations.

In this article, we experimentally explore and spatially resolve the skyrmion energy landscape of a magnetic thin film multilayer system. Utilizing Thiele model simulations, we develop a combined experimental and simulation approach that allows us to reconstruct the full, two-dimensional effective energy landscape governing thermal and driven skyrmion motion. This procedure also allows us to determine the conversion factor between experimental and simulation timescales, enabling fully quantitative simulations of dynamical states of hundreds or even thousands of μm-sized skyrmions. We show in our analysis that the method provides excellent descriptive simulation results even if the energy landscape has to be reconstructed under adverse experimental and statistical circumstances. Finally, we demonstrate the predictive power of our model for future applications by comparing simulation and experimental results for the so far experimentally unexplored relationship between skyrmion density and diffusion, whose behavior cannot be captured quantitatively by simulations ignoring pinning effects[43].

## Results

### Initial determination of the skyrmion pinning landscape

To observe skyrmion hopping dynamics and the form in which they are impacted by an inhomogeneous energy landscape, we utilize a Ta(5)/Co$_{20}$Fe$_{60}$B$_{20}$(0.9)/Ta(0.09)/MgO(2)/Ta(5) thin film multilayer stack (thickness of the layers in nanometers in parenthesis) to stabilize thermally active skyrmions at a temperature of 316 K[10]. Skyrmions are observed using a magneto-optical Kerr microscope in a 200 µm x 200 µm structure, in which the skyrmion density remains constant (see Methods for details). For the measurements, we choose a skyrmion density that is sufficiently low to minimize interactions between them, yet high enough to ensure adequate statistical data collection during the measurement period. The measured area is chosen to be away from the device boundaries to minimize skyrmion-boundary interactions[39]. Skyrmions at a low density (0.0080±0.0004 skyrmion/µm$^2$) are observed for two 1-hour-long measurements to determine the spatially resolved probability distribution as discussed in the methods section.

To determine the pinning potential from experimental measurements, the probability density $\rho(x,y)$ to find a skyrmion at a certain sample position $(x,y)$ can be used. The probability density is, in general, given by the Boltzmann weight

$$\rho(x,y) \propto \exp\left(-\frac{U(x,y)}{k_\mathrm{B}T}\right)$$

where $U(x,y)$ is the effective energy landscape and $k_\mathrm{B}T$ is the Boltzmann constant. The pinning energy landscape can therefore be calculated from a time-averaged experimental skyrmion distribution $\rho(x,y)$ by multiplying its negative logarithm with $k_\mathrm{B}T$.

Within the resulting pinning potential depicted in Fig. 1(a), some specific characteristics can be observed. First, during the measurement time, skyrmions are typically located at specific pinning sites with a very high probability, while on a large part of the sample (around 86 % of bins), no skyrmions are observed during the entire measurement time frame. These spaces remain indeterminate in the potential map (shown as white in Fig. 1(a)). To consistently sample all those areas, significantly longer and consequently experimentally difficult measurements would be required, the duration of which will be estimated later. Processes observed in experiments usually involve hopping motion with transitions through the unsampled areas, which cannot be dynamically resolved due to the limited time resolution of the Kerr microscope. Conversely, skyrmions within pinning sites typically stay pinned for many frames, leading to well-sampled regions. When two or more pinning sites with a low energy barrier between them are clustered together, the regions between sites can also be sampled due to the frequent transitions between them (see Fig. 1a, b). This allows for accurate simulations of skyrmion dynamics in these regions as long as the skyrmion stays within one pinning cluster. By labeling clusters and sites (the exact labeling process is described in the Methods section), we can assign all observed skyrmions to a specific pinning cluster and site based on their position. The process of labeling an individual trajectory is shown in Fig. 1(b) with the resulting labeling shown in Fig. 1(c). This labeled trajectory can then be employed to quantify the skyrmion hopping behavior and obtain the simulation timescale. For its determination, we accordingly focus on pinning sites, high probability regions, and clusters of multiple pinning sites within close proximity (see Fig. 1a,b). While we can directly infer that the energy between clusters is still mostly within the range of thermal excitations, the exact barrier height cannot be estimated directly from the measured skyrmion localization probability density. To obtain a complete description of the effective potential everywhere, we therefore develop a method to interpolate into these areas using simulations, which is described in the later sections.

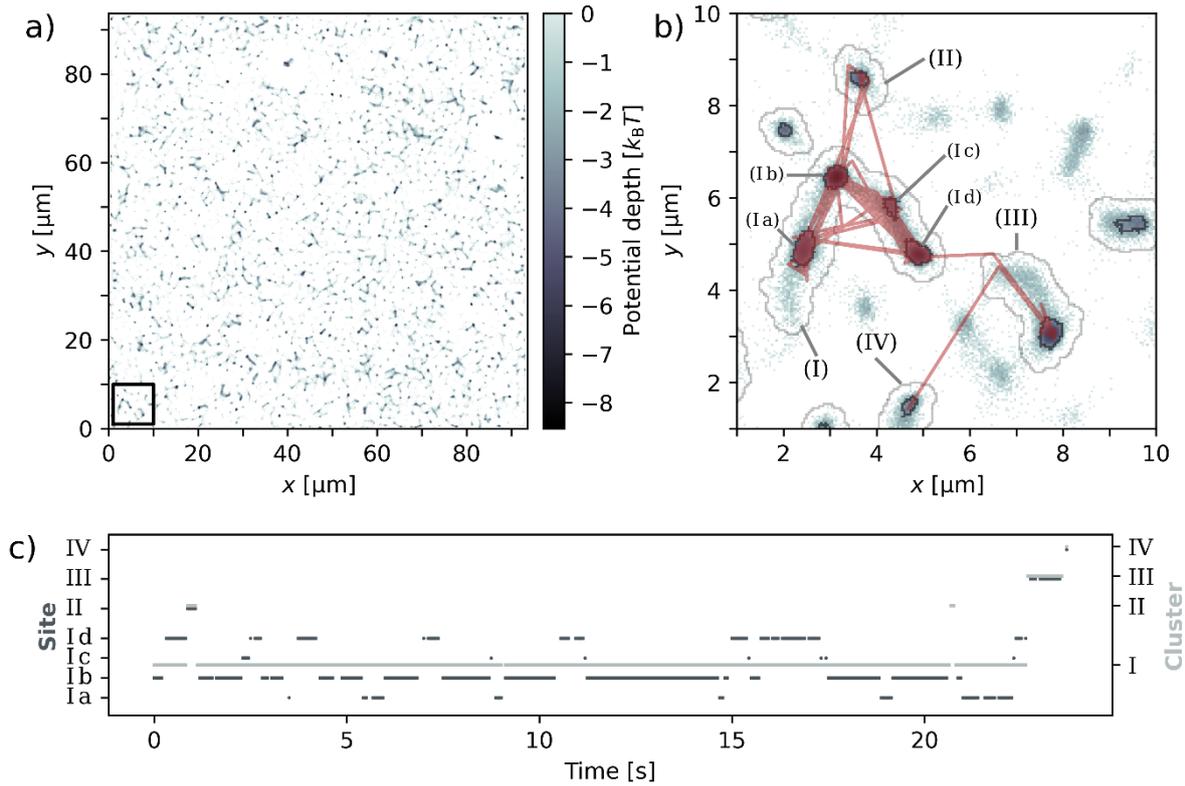

**Figure 1**: **Skyrmion pinning potential and hopping dynamics**. (a) Energy landscape quantified by the potential of mean force $U_{\text{pin}}(x, y)$ determined from occurrences of a skyrmion within the respective pinning potential bin. Areas in which no skyrmion is observed during the entire measurement time are color-coded as white. The color bar is identical for both (a) and (b). (b) shows a magnified part of the energy landscape (marked as a small black box in (a)). Clusters and sites are outlined in grey and black, respectively, and arbitrarily labeled with roman numerals (clusters) and letters(sites) to later identify the cluster and site the skyrmion occupies during each frame. The trajectory shown in red illustrates the typical hopping motion between sites where transitions are fast compared to the time spent within the pinning sites, with the transition to another site often happening within a single frame of the experimentally acquired video. The trajectory also shows a similar hopping between clusters with a significantly higher dwell time. (c) Site (dark blue) and cluster (grey), a particular skyrmion was located in for each frame. The skyrmion randomly switches between sites due to thermal excitations, mostly within the same cluster, due to the lower pinning energy barrier between the sites within a cluster. As the time for which the skyrmion stays pinned depends on the time it takes for a thermal excitation to surpass

the potential barrier out of the site, the dwell time is intrinsically random. Characteristic dwell times are related to the depth of the potential at the pinning site when compared to the surrounding potential.

## Coarse-grained simulations of magnetic skyrmions

Building on this phenomenological understanding of skyrmion dynamics, we now focus on their theoretical description in coarse-grained simulations. Skyrmion dynamics in the low-energy regime with no deformation of the spin structure can be described in good approximation by the Thiele equation[38,44,45] using a quasiparticle approach. It models the skyrmion as a quasiparticle with an intrinsic gyrotropic force term and is given as[44,46]

$$-\gamma \vec{v} - G_{\mathrm{rel}} \gamma \vec{z} \times \vec{v} + \vec{F}_{\mathrm{therm}} - \vec{\nabla} U_{\mathrm{pin}}(\vec{r}) + \vec{F}_{\mathrm{SkSk}}(\{\vec{r}\}) + \sum \vec{F}_{\mathrm{det,other}}(\vec{r}) = 0.$$

Here, the first two terms describe the skyrmion's response to the acting forces, which constitute the remaining terms. The thermal random force $\vec{F}_{\mathrm{therm}}$ is assumed to be Gaussian white noise. Its respective strength is given by the fluctuation-dissipation-theorem[47]. The interaction forces of skyrmions with each other $\vec{F}_{\mathrm{SkSk}}(\{\vec{r}\})$ are determined from experiment without prior assumptions about the functional form by applying the iterative Boltzmann inversion procedure described in Ref.[39] to this specific system. Additionally, $\vec{\nabla} U_{\mathrm{pin}}(\vec{r})$ models the pinning force as resulting from a pinning potential acting on the skyrmions, which is the subject of this work, and $\sum \vec{F}_{\mathrm{det,other}}(\vec{r})$ describes all other deterministic forces such as current induced forces[30], which are not considered in this study. Within those, $\vec{r}$ and $\vec{v}$ are the position and the velocity of the skyrmion under consideration, $\{\vec{r}\}$ represents the positions of all skyrmions, and $\gamma$ is the effective damping constant of skyrmion motion. This damping is related to the Gilbert damping $\alpha$ and diagonal entries of the isotropic skyrmion dissipation tensor $D_{\mathrm{diag}}$ via $D_{\mathrm{diag}} = \gamma/\alpha$. The skyrmion Hall effect, intrinsically linked to the skyrmion's non-zero winding number, is described by the relative gyrotropic force strength and can be determined from the effective skyrmion Hall angle $\theta_{\mathrm{eSH}}$ using $G_{\mathrm{rel}} = \tan(\theta_{\mathrm{eSH}})$. This angle is usually very small in our systems, where the stray field plays a dominating role in stabilizing µm-sized skyrmions with narrow domain walls[6,8,48,49], and therefore we can assume it to be zero.

To perform quantitative simulations using the Thiele approach, precisely determining all forces given within the Thiele equation is necessary. This means that the pinning potential (out of which the pinning force can be calculated) must be known with a spatial resolution that is sufficient for the potential to be considered smooth in the sense that the resulting forces do not exhibit significant discretization artifacts (examples, see Supplementary Material). The pinning potential must also have a defined value at every position in the simulation system. Thus, simulations can either be performed within only the sampled regions, or the missing potential regions need to be inter- or extrapolated. Additionally, the damping must be determined from experiments to yield quantitative comparisons. As damping and velocity only appear as a product, this is equivalent to a missing absolute timescale in simulations. Therefore, when applying standard simulation units $\gamma = 1$, $k_\mathrm{B} T = 1$, all time-like quantities will be given in units $\tau = \gamma(1\,\mathrm{\mu m}^2/k_\mathrm{B} T) = 1$ STU (simulation time unit). Hence, by determining the time conversion factor between simulation units and experimental units, the damping can be directly inferred.

## Obtaining simulation timescales from measured pinning site dwell times

To quantitatively model the dynamics of the skyrmion system within the Thiele model approach, the system damping (or, equivalently, the conversion from simulation to experimental time) must be ascertained. To obtain this, we analyze the distribution of dwell times, i.e., the time a skyrmion stays within a pinning site before hopping to another. For that purpose, we compare the distribution for the transitions between sites in well-sampled regions containing multiple pinning sites (which we refer to as pinning clusters) between simulation and experiment. The missing parts of the potential arising from limited statistics can here be mitigated as the skyrmion predominantly moves within the well-sampled regions of the potential, even when moving between sites. An exemplary cluster is shown in Fig. 2(c).

As the frame rate of the camera necessarily discretizes the experimental distribution, the high time resolution simulation trajectories are subsampled by only considering every $n^{\text{th}}$ step, where $n$ gives the conversion factor between simulation writeout time and experimental frametime which is determined iteratively as described below. This yields a set of distributions for a span of possible time conversion factors from which the optimal conversion parameter will be determined (more details in Methods). At a fixed time resolution (in this case, the camera frame rate), the distributions are well described by an exponential

$$P(t) = N \exp\left(-\frac{t}{\tau_{\text{esc}}}\right)$$

with the normalization constant $N = e^{1/\tau_{\text{esc}}} - 1$, the characteristic escape time of the site $\tau_{\text{esc}}$ and the dwell time $t$, with both times given as the number of frames between transitions. The distributions for both transition directions for an exemplary pair of sites are shown in Fig. 2(a) and 2(d), respectively. We limit our analysis to clusters with internal transitions and sufficient experimental statistics. Furthermore, we only consider sites with an observed escape time of $\tau_{\text{esc}} > 4$ frames (0.25 s). Faster transitions will have a high proportion of dwell times faster than the exposure time of the microscope. These dwell times, therefore, cannot be resolved in the experiment, as the described effects lead to significant statistical fluctuations. To determine the timescale, distributions for both experiment and simulation are fitted with an exponential, and the time conversion factor that results in the closest matching $\tau_{\text{esc}}$ between simulation and experiment is chosen for each site. The determined value of $\tau_{\text{esc}}$ as a function of the time conversion factor, along with the experimental value, is shown in Fig 2(b) and 2(e) for the two respective transition directions for the exemplary site. The overall timescale is then determined by taking the average of many observed transitions within multiple clusters fulfilling the aforementioned criteria, yielding 1 frame = $(1.34 \pm 0.10)$ STU, which corresponds to a damping of $\gamma = (0.047 \pm 0.004)\ (k_\text{B}T/\mu\text{m})/(\mu\text{m/s})$.

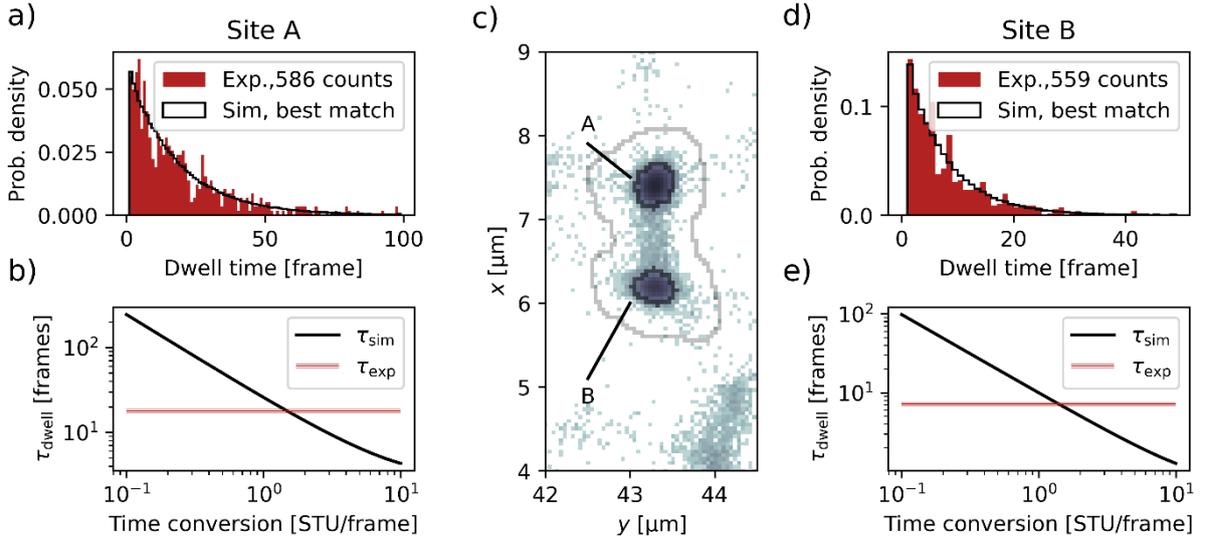

**Figure 2**: Determination of the time conversion factor between simulation and experiment for an exemplary pair of pinning sites. (a) and (d) show the experimentally determined dwell time distributions for the transitions $A \to B$ and $B \to A$, respectively, along with the simulation distribution at the time conversion factor best matching the two distributions. (b) and (e) show the characteristic dwell time in simulations as a function of the time conversion factor along with the given experimental value for the specific site. (c) shows the potential in the vicinity of the two sites with the two sites indicated by dark outlines and the boundary of the corresponding cluster shown via the grey outline.

### Reconstructing the effective pinning energy landscape to enable quantitative simulations

Large areas of the energy landscape cannot be effectively sampled due to the extremely long measurement time needed as the probability depends exponentially on the energy. Therefore, the potential of the unsampled areas of the pinning map still needs to be determined to achieve a complete description of the system's energy landscape. As skyrmions hop between clusters very quickly, the main effect of the unsampled region on the skyrmion dynamics is acting as an energy barrier. We approximate this barrier behavior by a simple, flat potential. While more complex extrapolation schemes could be applied using the same approach, with the specific version chosen here, no further assumptions about the functional form of the potential are needed. To find the effective barrier height, we run a set of simulations with barrier heights ranging from $0\ k_\mathrm{B}T$ to 6 $k_\mathrm{B}T$. We determine the diffusion by calculating the mean squared displacement (MSD) and fitting the linear region from 5 s to 15 s. This is the timescale at which transitions between clusters are common and linking with high confidence is possible. To account for the effect of trajectory linking on the diffusion, we run the same linking procedure on the simulation trajectories as is used for the experimental data acquisition process. We then fit an exponential to the diffusion constant as a function of the inserted barrier height and determine the intersection with the diffusion determined within the experiment (see Fig. 3). To calculate the experimental diffusion, we split the complete 2 hours of measurement into 10-minute segments and determine the diffusion for each segment and determine the mean along with the standard error of the mean. These intersections then give the energy barrier between clusters as $(2.78 \pm 0.03)\ k_\mathrm{B}T$. The level-dependent diffusion along with the relevant fits is shown in Fig. 3.

Using this determined fill level for the unsampled areas, we can now estimate the minimal measurement time needed to fully sample the energy landscape. The fill level corresponds to $\frac{e^{-\beta \cdot 2.78 k_\mathrm{B}T}}{2\ \text{hours}} = 0.031$ counts/hour. To reach a relative counting error of 10 %, 100 counts would be needed in every bin. This would result in a minimal continuous measurement duration of 134 days. Realistically, even more measurement time could be required, as this calculation assumes the best case of a flat unsampled region without anti-pinning sites, which would take exponentially longer to sample.

Additionally, the timescale allows for a direct comparison between simulated time and run time. The effective speed of the simulations performed in this work is within the order of magnitude of real time for even the high density (and therefore high skyrmion count) systems and even faster than real time for low density systems on current desktop hardware (details see Sup. Mat. Fig. 1). This makes running large parameter search simulations viable, enabling orders of magnitude faster and cheaper device design than experimentally possible.

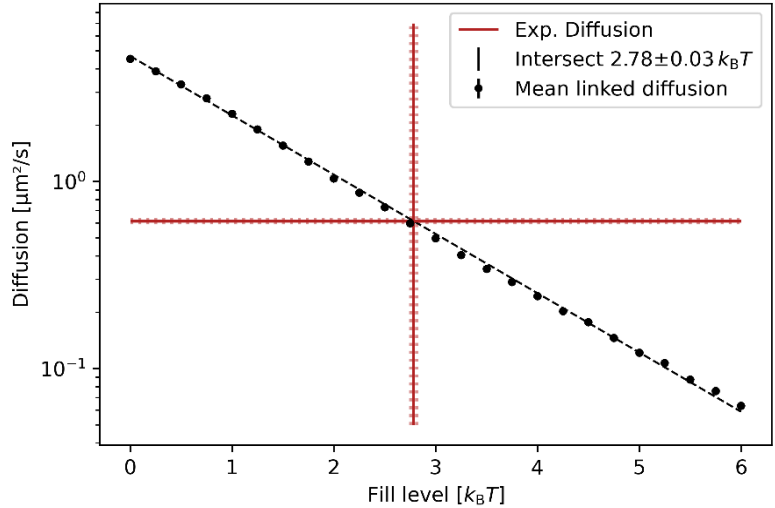

**Figure 3**: Determination of the fill level needed to match the experiment. The diffusion as a function of the fill level is shown for the trajectories that were linked using the Crocker-Grier linking algorithm[50] (black dots). The diffusion is then fitted by an exponential (dashed black line). To determine the experimental diffusion, the dataset is split into 10-minute segments and the mean and its uncertainty are determined (red horizontal line). The fill level of the intersect (red vertical line, error given as dashed lines) gives the confidence interval of the matched fill level for the final energy landscape of $(2.78 \pm 0.03)\, k_\mathrm{B}T$.

## Prediction and experimental validation of the relationship between skyrmion density and diffusion

To demonstrate the quantitative predictive power of this modeling approach, we will study the dependence of skyrmion diffusivity on the skyrmion density. The tradeoff between higher system complexity due to higher skyrmion number and lower diffusion due to the resulting density is an important consideration for applications such as reservoir computing[19,51]. The dependence of skyrmion diffusion on density has so far been studied only in simulations[43], but without considering realistic energy landscapes that include unavoidable pinning effects. An experimental study of such a system has been completely lacking, and quantitative simulation predictions for realistic experimental scenarios remained elusive. Within the system studied here as well as in previously studied systems showing skyrmion diffusion and varying skyrmion density[10,35,39], quantitative predictions of dynamics using micromagnetic or atomistic simulations are infeasible, as the system size and diffusion time far exceed computational limitations.

The experimental system studied is the same as above and can be seen for exemplary system configurations in Fig. 4(a). Here, different skyrmion densities are nucleated and the dependence of the diffusion coefficient on the density was evaluated. Figure 4(b) shows the experimentally measured diffusion constants along with the simulations results. The diffusion constant exhibits a clear decrease with increasing density, falling by two orders of magnitude within the measured density range. Simulation and experiment are compatible with the run-to-run variance of the simulation results for runs with the same duration. The large variance can be attributed to the inhomogeneous energy landscape as not all areas will be equally explored within 10-minute runs, leading to a spread much larger than would be expected from the statistical error of the diffusion determination. This agreement holds not only at the density at which the landscape was determined ($\approx 0.0080$ μm$^{-2}$) but extends to much higher and lower densities as well. To further the quantitative validation of our approach, the distribution of skyrmion displacements within 1 s is shown in Fig. 4(c) to (f). Both the one-dimensional and the two-dimensional distributions show good quantitative agreement between simulation and experiment, even for densities far away from where the effective energy landscape was recorded and on smaller timescales than those used for the diffusion matching process. As the diffusion constant is monotonically decreasing with skyrmion density, we can furthermore confirm our previous assumption that the skyrmion hall angle is negligible[43]. Moreover, our results reinforce the notion that a reliable modeling of pinning effects is key: predicting the diffusion without considering pinning effects leads to a low-density diffusion coefficient (($11.9 \pm 0.9$) μm$^2$s$^{-1}$) that differs from the experimental value by more than an order of magnitude!

Overall, we show that the simulations using the extracted timescale and energy landscape exhibit excellent quantitative agreement with experiments, even for strong variations in the experimental system, as long as those variations do not impact the pinning landscape. In particular, using our developed approach, all parameters can be determined directly from standard Kerr observation experiments within a few hours of experimental effort. This lifts the previous limitation of Thiele model simulations to qualitative predictions[34,41] and allows for a fully quantitative treatment of pinning energetics and two-dimensional diffusive skyrmion dynamics on a per-system basis.

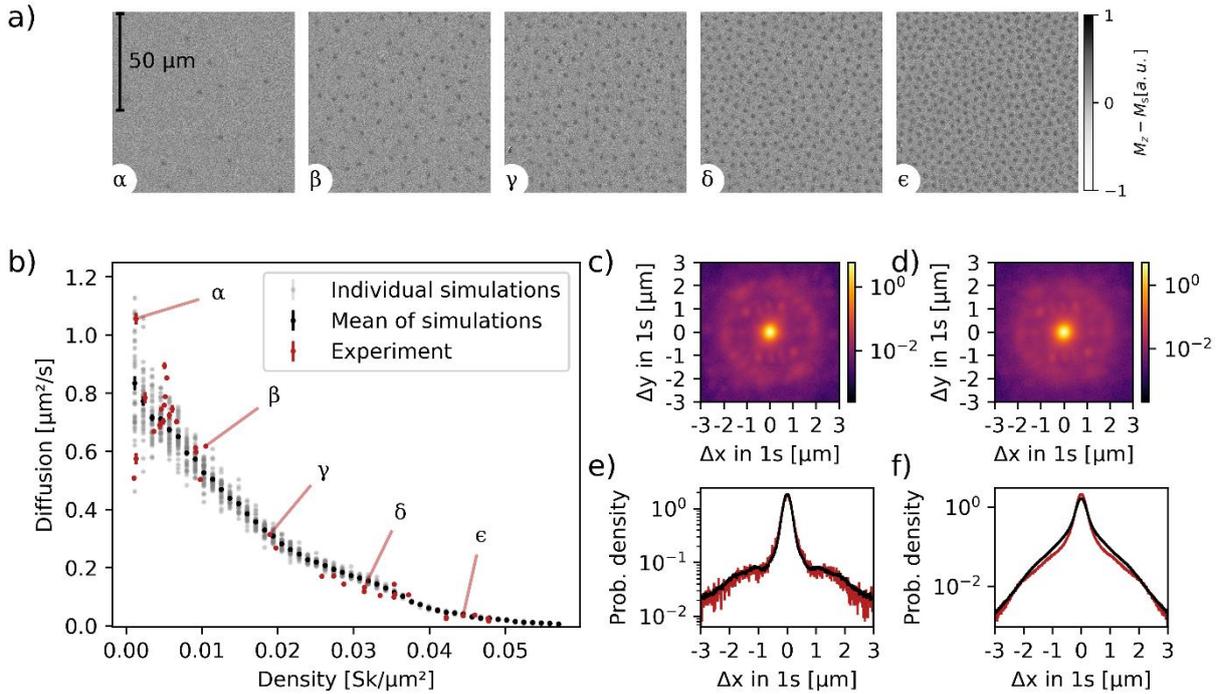

**Figure 4**: Experimental observation and simulation of the dependence of the diffusion coefficient on the skyrmion density. (a) Snapshot Kerr images of the measurements at different densities, marked with Greek letters. (b) Experimental (red) and simulated (black) diffusion coefficient data, using the unit conversion and potential of the unsampled areas determined in the sections before. For the simulation, both individual runs as well as the average for each density are shown. Experiment as well as simulation use individual runs of 10 minutes (c) Experimental and (d) simulation distribution of two-dimensional distance traveled by skyrmions over a time of 1 s at the matched density. The central peak corresponds to movement within one fixed pinning site. Secondary peaks are fixed by certain in-cluster transitions that dominate. (e) and (f) show the x displacement within 1 s for both experiment and simulation at densities α (e) and ϵ (f).

## Discussion

Here, we develop and experimentally demonstrate an easy-to-use method to obtain a two-dimensional effective energy landscape and quantitatively model pinned skyrmion dynamics of a real experimental sample. To enable realistic modelling, we determine the timescale conversion between experiment and simulation based on experimental hopping dynamics. With the effective pinning energetics and timescale conversion, we demonstrate the predictive power of our approach by determining the density dependence of pinned skyrmion diffusion with simulations which is in excellent agreement with corresponding experimental results. Thereby, we show that this ansatz enables quantitative simulations using the Thiele model on experimentally relevant scales and in real experimental units, addressing a critical gap in previous methodologies. Our method facilitates fully predictive simulations of experimental systems over large spatial and temporal scales, which are essential for applications such as skyrmion lattice formation[2,3,33,34] or device design[14,16,17]. Our method remains robust even for systems with low statistics and strong pinning. The only requirements for its application are the observation of hopping motion in most skyrmions and a few well-sampled transitions between distinct sites. As a result, it can also be extended to other particle-based systems, such as colloids, which can exhibit similar dynamics[52–56].

Accurately modeling the effects of an inhomogeneous energy landscape is particularly important in systems where thermal effects, pinning, and skyrmion-skyrmion interactions act on comparable scales, as small variations in energy can significantly influence system dynamics. These specifically include systems typically used for skyrmion-based non-conventional computing, such as Brownian computing[10,14,16–19] and reservoir computing[14,16,19,51]. For these unconventional computing paradigms, it is crucial to use a simulation approach fast enough to screen various potential designs at high throughput before experimental implementation as well as modeling many-particle effects such as the density dependent diffusion reduction[16,17,57]. While micromagnetic and atomistic approaches cannot provide this with currently available computational resources, our quasiparticle approach can simulate skyrmion dynamics in real time (as demonstrated in Supplementary Material Figure 2) and, using the findings presented here, provides quantitative results. It allows for the simulation of much larger system sizes and timescales, orders of magnitude beyond what is possible using micromagnetic or atomistic approaches. This is especially relevant given the growing interest in large systems such as skyrmion lattices and their dynamics[32,58,59]. This work thus provides a pathway to predictive in-silico modelling of skyrmion-based devices, which can be designed and tested by parallel simulations to identify interesting targets for more involved and time-consuming experiments.

## Acknowledgments

We acknowledge financial support from the Deutsche Forschungsgemeinschaft (DFG, German Research Foundation): Project number 403502522-SPP 2137 Skyrmionics. The authors are grateful for funding from TopDyn, SFB TRR 173 Spin+X (project A01, B02, A12, and B12 #268565370, TRR 146 (project #233630050), ERC-2019-SyG no. 856538 (3D MAGiC), and the Horizon Europe project no. 101070290 (NIMFEIA). The authors further acknowledge the computing time granted on the supercomputer MOGON II and III at Johannes Gutenberg University Mainz as part of NHR South-West.  M.K. acknowledges support by the Research Council of Norway through its Center of Excellence 262633 "QuSpin". M.A.B. was supported by a doctoral scholarship from the Studienstiftung des deutschen Volkes.

## Author Contributions

S.M.F. and T.S. jointly developed the timescale with the help of M.A.B. The level-matching procedure was developed by S.M.F., T.S., and M.A.B.. S.M.F. performed the simulations with the help of M.A.B. and J.R.; T.S. prepared the measurement setup and performed the Kerr microscopy measurements; F.K. and K.R. fabricated the multilayer sample. T.S. and S.M.F. evaluated the experimental and simulation data and prepared the manuscript supported by M.A.B., S.K., P.V., and M.K.. J.R., M.A.B., and S.M.F. provided the theoretical calculations. The study was supervised by M.K. and P.V.; all authors commented on the manuscript.

## Methods

### Thiele model simulations

The Thiele equation of motion[40,45] is integrated using the Heun integrator with a time step of $dt = 10^{-4}$. We employ standard parameters for numerical stability $\gamma_{\text{sim}} = 1$, $k_B T_{\text{sim}} = 1$, and convert to experimental units afterwards, if necessary, as discussed in the results section. Pinning forces exerted on the skyrmions are calculated from the binned pinning potential using finite differences to estimate the derivative. The skyrmion-skyrmion interaction[39] uses a truncated, shifted exponential potential derived from iterative Boltzmann inversion given by $U(r) = Ae^{\frac{-r}{B}} - U_c$ with parameters $A = 9623.9\ k_B T$, $B = 0.537\ \mu m$ and $U_c = 0.03\ k_B T = U(r_{\text{cut}} = 6.8\ \mu m)$. Since the rectangular geometrical structure effectively constrains the skyrmion dynamics, but the edges of the structure are far outside the experimental field of view, we perform all simulations using periodic boundary conditions. The dwell time distribution in simulations was found using a simulation of 1000 non-interacting particles with a combined trajectory duration of $1 \times 10^9$ STU for each cluster. Level matching was performed using 24 independent simulations of 70 interacting particles with a combined trajectory duration of $8.75 \times 10^7$ STU per fill level with fill levels ranging from $0\ k_B T$ to $6\ k_B T$ in steps of $0.25\ k_B T$. Simulations for the density-dependent diffusion were run for a time corresponding to 10 minutes per density. To compensate for the lower statistics due to the decreased particle number at lower densities, 40 simulations were run for the lowest 10 densities, 30 for the next 10 higher densities, 20 for the next 10 and 10 for each of the higher densities. Simulation trajectories for the level matching and density-dependent diffusion were performed at a writeout frequency of 1 frame/writeout as determined by the timescale matching.

### Sample characterization and structure fabrication

The thin film layer stack used was sputtered by a Singulus Rotaris magnetron sputtering system at a base pressure of 3×10$^{-8}$ mbar. It consists of Ta(5)/Co$_{20}$Fe$_{60}$B$_{20}$(0.9)/Ta(0.09)/Ta(5)/MgO(2) with the thickness of the layers in nanometers in parenthesis. The subscript numbers represent the percentual atomic concentration. Our setup allows the layer thicknesses to be reproducible with an accuracy better than 0.01 nm[10].

Electron beam lithography (EBL) was then employed to pattern the structures using a Raith Pioneer Electron Beam system and subsequent Argon ion etching using an IonSys Model 500 ion beam etching system with endpoint detection. The samples were structured so that specific parts of the energy landscape could be found again to be revisited and to guarantee an approximately constant skyrmion density for the measurement duration.

## Measurement setup and skyrmion nucleation

For imaging, we used the polar magneto-optical Kerr effect (MOKE) in a commercially available Kerr microscope from *evico magnetics GmbH*. Perpendicularly and parallelly aligned electromagnetic coils allow for applying in-plane (IP) and out-of-plane (OOP) magnetic fields simultaneously. The OOP magnet is custom-made and specifically constructed to provide small fields with sub-microtesla precision. The sample temperature was regulated using a Peltier element directly on top of the magnetic coil, and a Pt-100 heat sensor was used to measure the temperature close to the sample with an accuracy of better than 0.2 K. All the measurements were performed at 316.0+/-0.2 K. The Kerr microscope and measurement setups are enclosed by a thermally stabilized flow box with a total thermal stability of 0.2 K. A CCD camera records gray-scale videos of the magnetic contrast with a field of view of 123 × 94 µm$^2$ at 16 frames per second, i.e., a time resolution of 62.5 ms.

The skyrmions are nucleated at a fixed OOP field by applying a large IP field pulse. Thereby, skyrmions are stabilized when the IP field is switched off. By adjusting the OOP field at which the skyrmions are nucleated, the nucleated skyrmion density can be controlled. Additionally, the applied OOP field allows us to directly tune the skyrmion size. Consequently, the number of skyrmions can also be tuned by increasing the OOP field to the point where the size of the skyrmions is sufficiently reduced for skyrmions to start annihilating. When keeping different skyrmion densities at a constant OOP field, for higher densities, due to the stray field energy, the skyrmion size is reduced. As the pinning changes with changing skyrmion size, and to allow for the Thiele model to hold for all densities, the skyrmion size was kept constant manually at a radius of 1.0+/-0.2 µm by adjusting the applied OOP field for different densities.

## Skyrmion tracking, determination of diffusion coefficients, and pinning cluster localization

We use a trained convolutional neural network to detect skyrmions in Kerr microscopy image data[50] to preprocess and detect the skyrmion positions from the Kerr microscopy gray-scale videos. It allows for the insertion of images of size 512 × 512 px and, therefore, only a part of the image of size 93.6 × 93.6 µm$^2$ is used. The network labels skyrmions pixel-wise, with the center of mass of the detected skyrmion is taken as the skyrmion position. As a large number of pixels is taken into account to determine the skyrmion position, which is then only represented as the center of mass, it can be determined with an accuracy higher than the optical resolution of the microscope. The contrast is enhanced by applying background subtraction with respect to a saturated state. Within a measurement duration of below one hour, the drift was found to be smaller than the bin resolution of the energy landscape. To correct for drift occurring between different new nucleation sets of skyrmions, the resulting effective pinning maps (before filling the bins with no counts) are overlapped, and the difference is minimized. The positions are then linked using trackpy's implementation of the Crocker-Grier linking algorithm[60,61]. To estimate the effect of the linking error, we use the same linking procedure on simulations and compare the resulting trajectories with the ground truth. This comparison was made for the full set of level matching simulations, showing the effect ranging from a reduction of ≈ 35% to an increase of ≈ 30%. To account for this effect, we perform the level matching by comparing the experimental diffusion to the diffusion determined from linked simulation trajectories. Without this correction, the determined fill level would be overestimated. Consequently, this would lead to an apparent match in diffusion coefficients between experimental and true simulation trajectories, even though the energy landscape would not accurately represent the underlying energy barriers. This allows for the creation of an "effective energy map" that incorporates the diffusion reduction, leading to matched diffusion between linked experimental and true simulation trajectories.

To determine the effective energy landscape, we bin the skyrmion occurrences with a fixed spatial resolution. For a sufficient smoothness, the potential discretization is chosen to be 45.7±0.5 nm, which corresponds to 4x4 bins within each experimental pixel to avoid pixelation artifacts and provide a good tradeoff between resolution and statistics.

As both single pinning sites and pinning clusters within this landscape are required for the purpose of this study, they must be well-defined. We define a pinning site as any area of directly connected bins where the calculated potential is below $-2.5\ k_\mathrm{B}T$ for all bins and which contains at least one bin with a potential below $-3.5\ k_\mathrm{B}T$ (all potential levels given as values before determination of the potential of the unsampled areas). The potential thresholds were chosen by considering clusters with multiple clearly separate sites (such as in Fig. 1(b)) and selecting a level sufficient to separate these sites while keeping areas without a clear boundary connected. This ensures pinning sites are clearly separated and well-sampled for all further analysis while still retaining a high number of well-defined pinning sites. Additionally, we define pinning clusters as connected regions containing at least one pinning site (thus requiring at least one bin within the region with a potential below $-3.5 k_\mathrm{B}T$) where all bins have a known potential. A simple definition of pinning clusters could be connected regions with at least one count in each bin. This, however, has to be adjusted to account for the limited frame rate in the experiment, as skyrmions often move by more than one bin within a single frame. As a result, the set of bins visited by a skyrmion, even when the skyrmion is pinned within a cluster, can become disconnected, especially in the cluster regions with poor statistics. To avoid labeling this behavior as leaving the cluster, the pinning clusters are expanded by 250 nm in all directions. Additionally, all clusters and sites connected to the edge of the field of view are excluded from further analysis, as leaving processes cannot be accurately determined for them.

To simulate and evaluate the dwell times for pinning sites within one cluster, a potential for the unsampled areas needs to be chosen at which the simulations can be performed. For the scope of this work, we choose it to be $4k_\mathrm{B}T$ to achieve sufficient separation from the pinning cluster potential but avoid discretization artifacts that can be introduced for large potential jumps between bins. From the resulting simulation and the measured experimental data, only skyrmion dynamics within pinning clusters, where the experimental potential is known everywhere, are considered, so the exact value chosen does not impact the resulting calculated intra-cluster dwell times.

To compare dwell time distributions between simulation and experiment, the simulation trajectory needs to be discretized according to the time conversion factor such that one writeout corresponds to one frame. For the determination of the timescale, this is done by running the simulations at a very high time resolution of 0.01 STU/writeout first and then down-sampling to lower effective writeout frequencies by taking every $n^\mathrm{th}$ writeout allowing different time conversion factors without rerunning the simulation. The time conversion factor is then $0.01n$ STU/frame. For the level matching, we simply set the simulation writeout interval to 1 frame, as the timescale is already known. The dwell time distributions are then compared by fitting them with an exponential of the form

$$P(t) = \left(e^{1/\tau_\mathrm{esc}} - 1\right) \exp\left(-\frac{t}{\tau_\mathrm{esc}}\right)$$

and comparing the characteristic dwell time $\tau_\mathrm{esc}$ determined from the fit. The probability density function is normalized such that $\sum_{t=1}^{t=\infty} P(t) = 1$ as the distribution is discrete and dwell times of 0 cannot occur.

Finally, we determine the skyrmion diffusion coefficients in experiment and simulation by fitting the linear part of the mean squared displacement that describes the hopping motion outside of a single pinning site. We exploit the relation $\langle [\Delta x(\Delta t)]^2 \rangle = 2dD \cdot \Delta t$, where $d = 2$ is the effective dimension and $\langle [\Delta x(\Delta t)]^2 \rangle$ is the mean squared displacement (MSD) during a time interval of $\Delta t$, and angled brackets indicate the average over all segments of skyrmion trajectories with length $\Delta t$ in time (using the sliding window method). To avoid sampling bias towards longer trajectories (that would contribute disproportionately due to the sliding window approach), the MSD is first calculated for each trajectory individually, and the overall MSD is determined from a weighted average with the weights given by the duration of the individual trajectories. The MSD is then fitted by a linear function in the interval from 5 to 15 seconds.

# Supplementary Material

## Pinning Potential Resolution

Competing effects constrain the choice of bin width for the potential: As the determination of the potential is heavily limited by the available experimental statistics, a comparably large bin size is preferred to increase the number of skyrmion observations per bin and reduce the number of bins without observations. However, this competes with the need for a high-resolution energy map to provide accurate force calculations for the simulation. We illustrate this effect for an exemplary, Gaussian-shaped pinning site of width $\sigma$ and depth $\epsilon$ in Sup. Mat. Fig. 1. The coarse sampling at $1\sigma$ shows clear discretization artifacts and does not recover the potential depth correctly. A finer sampling at $0.25\sigma$ gives a much better approximation of the force and potential as well as a much smaller reduction in the potential depth compared to the original Gaussian. For low sampling point density, the numerical forces calculated from the potential systematically underestimate the potential depth and misrepresent the site shape and effect. In simulations, this has a significant impact on skyrmion dynamics, as the site depth is of critical importance for both the timescale and level-matching methods demonstrated in this work. We therefore need to choose a bin width that is as fine as possible while still keeping the number of unsampled bins as low as possible.

This avoids possible issues due to subpixel-induced artifacts. As the pinning sites in our system (when fitted with Gaussians) typically have a width $\sigma$ on the order of $200 - 400$ nm, we choose a bin width of 45.7 nm (1/4 of the width of a pixel) as this will result in a force approximation that provides for smooth dynamics even when considering small pinning sites within our system (comparable to or better than the fine sampling shown in Sup. Mat. Fig. 1).

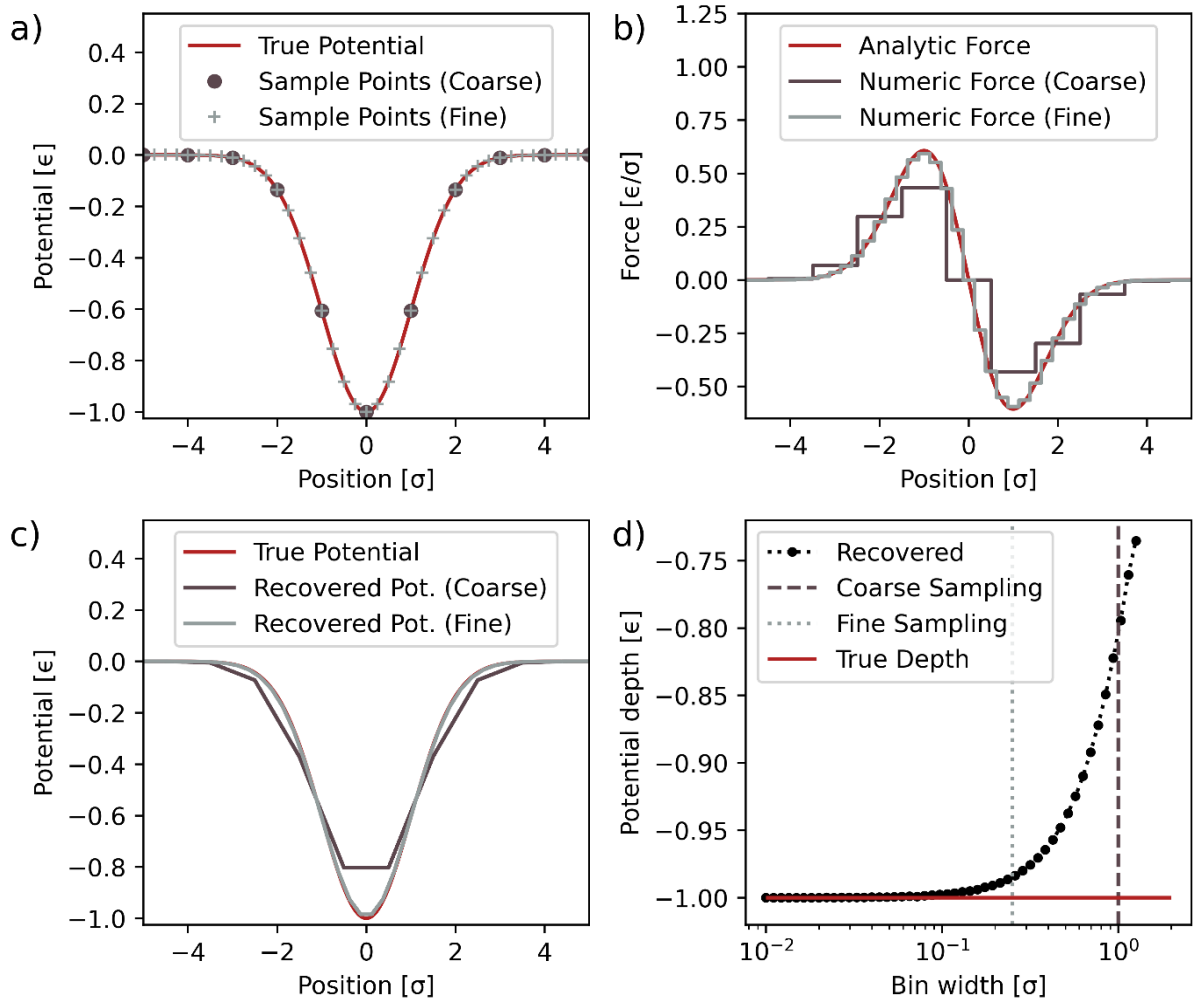

Supplementary Material Figure 1: Illustration of the effect of binning resolution on the resulting numerical force and recovered potential. a) shows an exemplary Gaussian potential along with two sampling resolutions, one coarse sampling at a spacing of $1\sigma$, shown in brown, and a finer sampling at $0.25\sigma$ (shown in grey, comparable to the resolution used in this work for small pinning sites). b) shows the force, calculated as a finite difference approximation of the derivative of the points at which the potential is sampled. Both samplings result in step functions because the force within a bin is always constant. The approximation approaches the analytic force for higher resolutions. The coarse sampling clearly underestimates the attractive force at the lowest point of the potential. c) shows the potential recovered by integrating the numerically derived forces, resulting in linear segments. While the recovered potential when implementing the fine binning is in good agreement with the true Gaussian potential, deviations in the coarse recovered potential are apparent, both in its shape and its total depth. d) shows the recovered potential depth as a function of the chosen sampling resolution. Finer bin widths approach the true depth asymptotically. The relative error at a bin width of $0.25\sigma$ (fine sampling) is 1.5 %.

## Simulation performance

Due to the varying complexity of the systems simulated in this work, performance needs to be evaluated on a per-system basis. Since we consider ensembles of many runs, we chose to trivially parallelize the simulation procedure by running one simulation per CPU thread and perform multiple simulations at the same time. The effective simulation performance on a single system is therefore the aggregate of all parallel runs. Using modern CPU hardware (Ryzen 9 7950X3D), the effective simulation speed ranged from $\approx 30$ times faster than real time for the lowest density system (10 skyrmions) to around 3 times slower than real time for the highest density (500 skyrmions) with real-time matching simulation time at around 260 skyrmions (shown in App. Fig. 1). This level of performance makes it possible to run large numbers of system configurations on a compute cluster to test device designs etc. to mature them much faster and less resource intensively than would be possible with manufacturing and measuring physical devices.

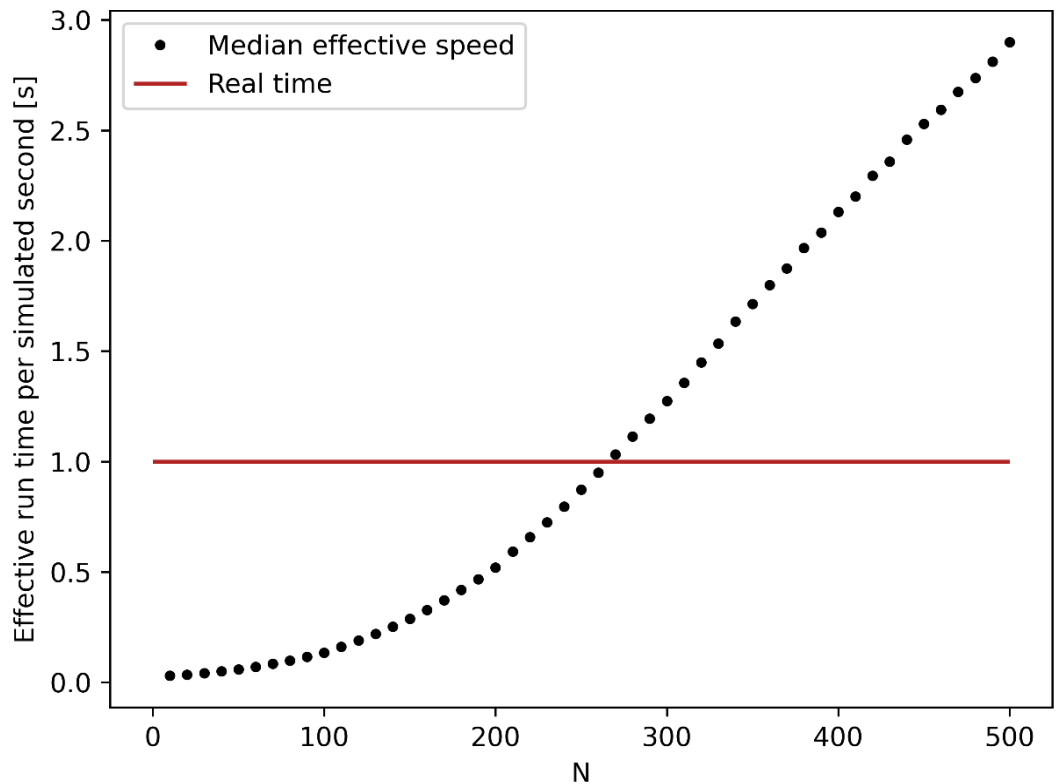

Supplementary Material Figure 2: Simulation performance on modern desktop hardware (AMD Ryzen 9 7950X3D). The effective simulated time (aggregated over parallel simulations) is shown for a range of different particle numbers $N$. Speed varies between around 30 times real time ($N = 10$) to around 1/3 real time ($N = 500$) with real time performance reached around $N = 260$. Performance data is taken from the density-dependent diffusion simulations used to show the model's predictive power (see Fig. 4). Faster individual runs occur when the CPU is no longer fully loaded at the end of batches of runs.